\documentclass{phb-proc4-auth}
\usepackage{graphicx}
\usepackage{amssymb}
\begin{document}
\begin{frontmatter}
\journal{SCES '04}

\title{Ferromagnetic fluctuation and superconductivity in Na$_{0.35}$CoO$_2 \cdot$1.3H$_2$O: FLEX study of multi-orbital Hubbard model}
\author{Masahito Mochizuki\corauthref{1},}
\author{Youichi Yanase}
\author{and}
\author{Masao Ogata}
\address{Department of Physics, University of Tokyo, Hongo Bunkyo-ku Tokyo 113-0033, Japan}
\corauth[1]{Corresponding Author: University of Tokyo, 7-3-1 Hongo Bunkyo-ku Tokyo 113-0033, Japan.
Phone: +81-3-5841-4185, Email: motiduki@hosi.phys.s.u-tokyo.ac.jp}

\begin{abstract}
Spin and charge fluctuations and superconductivity in Na$_{0.35}$CoO$_2\cdot$1.3H$_2$O are studied based on a multi-orbital Hubbard model. By applying the fluctuation exchange (FLEX) approximation, we show that the Hund's-rule coupling between the Co $t_{2g}$ orbitals causes ferromagnetic spin fluctuation. Triplet pairing is favored by this ferromagnetic fluctuation on the hole-pocket band. We propose that, in Na$_{0.35}$CoO$_2\cdot$1.3H$_2$O, Co $t_{2g}$ orbitals and inter-orbital Hund's-rule coupling play important roles on the triplet pairing, and this compound can be a first example of the triplet superconductor mediated by inter-orbital-interaction-induced ferromagnetic fluctuation.
\end{abstract}

\begin{keyword}
Na$_{0.35}$CoO$_2\cdot$1.3H$_2$O \sep orbital degree of freedom  \sep Hund's-rule coupling  \sep triplet superconductivity
\end{keyword}
\end{frontmatter}

\begin{figure}[tdp]
\includegraphics[scale=0.6]{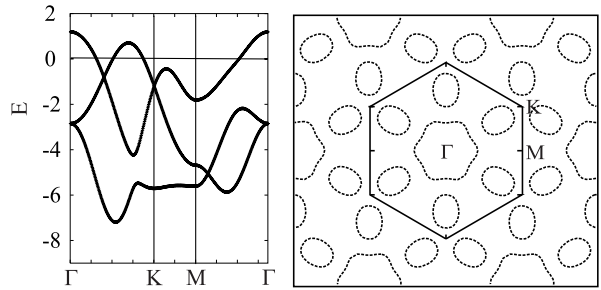}
\caption{Band dispersions (left panel) and Fermi surfaces (right panel) obtained from $H_{\rm kin.}+H_{\rm cry.}$. Both reproduce very well the characters of the LDA results (Ref.~\cite{Singh00}) especially near the Fermi level.}
\label{FermiS}
\end{figure}

The mechanism and nature of superconductivity in Na$_{0.35}$CoO$_2 \cdot$1.3H$_2$O is currently attracting great interest~\cite{Takada03}. However, in spite of a lot of attempt and effort, even the paring symmetry has not been resolved yet. In this paper, we study the electronic structure and superconductivity in this compound by employing a multi-orbital Hamiltonian given by $H=H_{\rm cry.}+H_{\rm kin.}+H_{\rm int.}$. Here, the first term $H_{\rm cry.}$ expresses the crystal field from the O ions acting on the Co $t_{2g}$ orbitals, and the second term $H_{\rm kin.}$ represents the kinetic energy. The tight-binding parameters are deduced by fitting the LDA band structure, and we find that up to the third nearest-neighbor hoppings between Co $t_{2g}$ orbitals are necessary at least. In Fig.~\ref{FermiS}, we show the band dispersions and the Fermi surfaces for non-interacting Hamiltonian ($H_{\rm cry.}+H_{\rm kin.}$). Both reproduce very well the LDA results~\cite{Singh00}, especially near the Fermi level. The Fermi surfaces consist of a large cylindrical one around the $\Gamma$-point and six hole pockets near the K-points. The cylindrical Fermi surface has a dominant $a_{1g}$-orbital character, while the hole pockets have an ${e'}_g$-orbital character.

The term $H_{\rm on-site}$ represents the on-site $d$-$d$ Coulomb interactions, which consists of following four contributions: $H_{\rm int.}=H_{U}+H_{U'}+H_{J_{\rm H}}+H_{J'}$ where $H_{U}$ and $H_{U'}$ are the intra- and inter-orbital Coulomb interactions, respectively, and $H_{J_{\rm H}}$ and $H_{J'}$ are the Hund's-rule coupling and the pair-hopping interactions, respectively. These interactions are expressed using Kanamori parameters, $U$, $U'$, $J_{\rm H}$ and $J'$. The value of $U$ has been estimated experimentally as 3-5.5 eV, and $J_{\rm H}$ for Co$^{3+}$ ion is 0.84 eV. Thus, the ratio $J_{\rm H}/U$, which gives the strength of Hund's-rule coupling, is 0.15-0.28 in this compound.

\begin{figure}[tdp]
\includegraphics[scale=0.45]{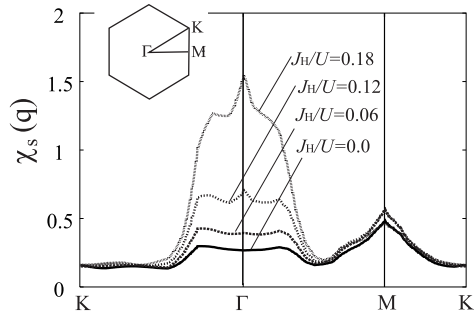}
\caption{Spin susceptibility showing the evolution of ferromagnetic fluctuation as the strength of Hund's-rule coupling $J_{\rm H}/U$ increases.}
\label{FMfluc}
\end{figure}
We apply the FLEX approximation to this multi-orbital Hubbard model. Figure~\ref{FMfluc} shows the spin susceptibility $\hat{\chi}^{\rm s}$ for various values of $J_{\rm H}/U$ at $T=$0.02. As shown in this figure, a peak structure around $\Gamma$-point is strongly enhanced as $J_{\rm H}/U$ increases, while the small peak at the M-point hardly changes. Since $J_{\rm H}/U$ is 0.15-0.28 in the actual compound, the enhanced peak at $\Gamma$-point at $J_{\rm H}/U=$0.18 indicates that the ferromagnetic fluctuation actually exists in this compound. Further, the structure at the $\Gamma$-point is considerably small without Hund's-rule coupling ($J_{\rm H}/U=0$), indicating a substantial importance of the Hund's-rule coupling for emergence of the ferromagnetic fluctuation. On the other hand, the charge-orbital susceptibility does not exhibit any remarkable structures, implying an absence of the charge- and orbital-density-wave instabilities.

\begin{figure}[tdp]
\includegraphics[scale=0.45]{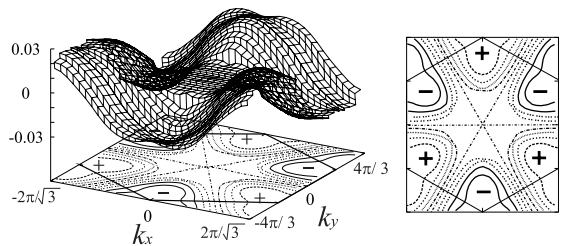}
\caption{$k$-dependence of the $f_{y(y^2-3x^2)}$-wave gap 
at $J_{\rm H}/U=0.18$.}
\label{fwaveSC}
\end{figure}
By solving the Eliashberg equations,  we find that the dominant superconducting instability at $J_{\rm H}/U=0.18$ is $f_{y(y^2-3x^2)}$-wave triplet pairing (see Fig.~\ref{fwaveSC}), and is generated on the ${e'}_g$ band with the pocket Fermi surfaces. On the other hand, the amplitude of gap on the $a_{1g}$ band with a large cylindrical Fermi surface is considerably small. This gap structure is consistent with several NQR/NMR experiments~\cite{Fujimoto03,Ishida03}, which suggest existence of line nodes. Further, absence of broken time-reversal symmetry observed in $\mu$SR~\cite{Higemoto03} is also consistent with our result.

The dominant contribution of the ${e'}_g$ band to the superconductivity is attributed to the van Hove singularity (vHS) in the ${e'}_g$ band. As can be seen in the band structure in Fig.~\ref{FermiS}, there exist saddle points near the K-points. This leads to a large density of states (DOS) near the Fermi level. 

In fact, the domination of $f_{y(y^2-3x^2)}$-wave pairing can be understood when we consider the geometry of Fermi surfaces. In this gap structure, the nodal lines run between the hole pockets, and do not intersect them. Consequently, the gap is fully opened on each pocket Fermi surface~\cite{Kuroki04a,Kuroki04b}.
Thus, we expect that in this compound, the hole pockets play a substantially important role on the emergence of superconductivity. In fact, we have also studied cases without hole pocket by tuning the transfer parameters, and have shown that the pairing instability is strongly suppressed unless the system has the hole pockets. We expect that the hole pockets would be observed in the superconducting hydrate compound if the ARPES or de Haas-van Alphen measurements could successfully be performed.

In summary, we have studied the spin and charge fluctuations and the superconductivity in Na$_{0.35}$CoO$_2 \cdot$1.3H$_2$O based on the FLEX analysis of the multi-orbital Hubbard model. We have shown that the Hund's-rule coupling between the Co $t_{2g}$ orbitals gives rise to the ferromagnetic spin fluctuation. The $f_{y(y^2-3x^2)}$-wave pairing instability mediated by this ferromagnetic fluctuation arises mainly on the ${e'}_g$ band with the pocket Fermi surfaces. The pocket Fermi surfaces as well as a large DOS due to the vHS near the Fermi level are crucially important for the $f_{y(y^2-3x^2)}$-wave pairing instability on the ${e'}_g$ band. We have pointed out that this material can be a first example of the inter-orbital-interaction-induced triplet superconductivity.

We thank K. Ishida, G.-q.Zheng, K. Yoshimura, K. Kobayashi, M. Sato, R. Kadono, K. Kuroki, G. Baskaran for valuable discussions. M.M also thanks S. Onari and Yasuhiro Tanaka for discussions. This work is supported by ``Grant-in-Aid for Scientific Research'' from the MEXT of Japan.

\noindent
{\it Note added. ---} 
After the submission of this paper, our further study found a tiplet $p$-wave state which is nearly degenerate with the $f$-wave one proposed in this paper. This $p$-wave state is also mediated by the FM fluctuation due to the Hund's-rule coupling, and the hole pockets and the vHS singularity also play substantial roles. For details, see cond-mat/0407094.

\end{document}